\documentclass[twocolumn,showpacs,preprintnumbers,amsmath,amssymb]{revtex4}

\usepackage{graphicx}
\usepackage{bm}

\begin{document}


\title{Effects of fermion-boson interaction in neutral atomic systems}

\author{Nobuhiko Yokoshi}
\author{Susumu Kurihara}%
\affiliation{%
Department of Physics, Waseda University, Okubo, Shinjuku, Tokyo 169-8555, Japan }%

\date{\today}

\begin{abstract}
We investigate the collective excitations of $^3$He-$^4$He mixture films at zero temperature within random phase approximation and linear response theory. In low concentration regime of $^3$He, a level repulsion between zero sound and third sound mode is derived, which opens the possibility to observe quantum mechanical coherence between $^3$He particle-hole pair and the condensate of $^4$He. We also investigate the $^3$He-$^4$He vertex corrections in ladder approximation and show that the third branch, the combined mode of fermionic particle-hole pair and the third sound quanta, provides a unique correction to Landau $f$ function. Some implications to the fermion-boson mixture of alkali atoms in a potential trap are discussed.
\end{abstract}

\maketitle


Two dimensional (2D) Fermi liquid is one of the most important subjects in condensed matter physics; they are in deep connection with high-temperature superconductivity in cuprates and alkali-doped fullerene, and with quantum Hall effect. Free surface of superfluid $^4$He bulk liquid or films provides ideal fields for 2D fermi systems free from impurities and inhomogeneities in chemical potential due to randomness of walls or substrates. In particular, phase separated $^3$He-$^4$He mixture films has been a subject of theoretical and experimental interest for the last two decades~\cite{rf:1}. At very low temperature, it is well-known that $^3$He atoms are bound to the surface of superfluid $^4$He~\cite{rf:2}, and behave as a well-defined 2D Fermi liquid~\cite{rf:3,rf:4}. Therefore, it is regarded as a good candidate of 2D fermion-boson systems. The nature of $^3$He and $^4$He films depends strongly on system parameters, such as concentrations, temperature, and van der Waals potential from substrates. This richness of the parameters has provided various suggestions such as Cooper pairing~\cite{rf:5} and dimerization~\cite{rf:6,rf:7,rf:8} of $^3$He, as well as suppression of the superfluidity~\cite{rf:9} and Casimir effect~\cite{rf:10,rf:11} of superfluid $^4$He films. 

In this 2D fermion-boson system, $^3$He particles interact with one another through thickness variation of $^4$He films, i.e., third sound driven by van der Waals potential~\cite{rf:5}, as well as direct interactions. A fascinating feature of this system is that characteristic energies can be easily tuned by varying the thickness of $^3$He and $^4$He films continuously. Therefore, it is expected that interference between excitations is enhanced, and nonadiabatic effect becomes relevant when suitable parameters are chosen. In this paper, we report a theoretical study of response to the spectrum of collective excitations when thickness of the mixture films is varied at zero temperature. We will show that a level repulsion between zero sound and third sound mode takes place. Moreover, the third branch is derived, which we interpret as combined mode of fermionic particle-hole pair and third sound phonon, by calculating the vertex function in some detail. We also calculate the contribution of this collective excitation to Landau $f$ function.

We assume that the system forms a phase-separated double layer, i.e., normal $^3$He liquid covering $^4$He which consist of a superfluid layer and a non-superfluid ``inert layer"~\cite{rf:9,rf:12}. In this system, particles are sensitive to the modulation of substrate van der Waals potential due to the thickness variation of $^3$He and $^4$He films, especially at low concentrations. We take account of this effect into interactions between particles and include other effects in hydrodynamic masses. The effective Hamiltonian which consists of third sound phonons and $^3$He quasiparticles interacting with one another and phonons was derived by one of the authors~\cite{rf:5}, and has the following form
\begin{eqnarray}
\!\!\!\!\!\!\!&H_{\rm{eff}}&=\sum_{\bf k,\sigma}\frac{\hbar^2 k^2}{2m_3}c_{\bf k,\sigma}^{\dagger}c_{\bf k,\sigma}+v_3\sum_{\bf{q}}\rho_{\bf q,\uparrow}\rho_{-\bf q,\downarrow}\nonumber
\\&+&\sum_{\bf q}\hbar\omega_{\bf q}b_{\bf q}^{\dagger}b_{\bf q}+\sum_{\bf q,\sigma}g_{\bf q}(b_{\bf q}+b_{-\bf q}^{\dagger})\rho_{-\bf q,\sigma},
\end{eqnarray}
where correlation between superfluid and inert layers and surface tension of the films are neglected for simplicity. These approximation may be allowed, except the region where the structure normal to substrate becomes remarkable~\cite{rf:13}. Here, $\rho_{{\bf q},\sigma}=\sum _{\bf k}c_{{\bf k},\sigma}^{\dagger}c_{\bf{k} + \bf{q},\sigma}$ is Fourier transformation of density operator and $\sigma=\uparrow ,\downarrow$ and $N_4$ are spin indices and number of $^4$He atoms respectively. The spectrum of third sound phonon and coupling energies are respectively given by
\begin{eqnarray}
(\hbar \omega_{\bf q})^2&=&(\hbar c_{\rm{B}}q)^2+\bigl(\frac{\hbar^2 q^2}{2m_4}\bigr)^2
\\v_3&=&\frac{3u_3}{2n_3^2(d+h_3+h_4)^4S}
\\g_{\bf q}&=&\frac{1}{\sqrt{N_4}}
\frac{3u_3h_4}{n_3(d+h_3+h_4)^4}
\bigl(\frac{\hbar^2q^2}{2m_4\hbar\omega_{\bf q}} \bigr)^{\frac{1}{2}} \,\,\,,
\end{eqnarray}
where the subscripts $3$ and $4$ denote $^3$He and $^4$He; $n_i$, $h_i$ are average bulk densities and average thickness of the films, and $m_i$, $u_i$ are hydrodynamic masses and characteristic van der Waals energies. $S$ and $d$ are surface area and thickness of inert layer. In the equations above, $c_{\rm{B}}$ is the velocity of third sound phonon which strongly depends on concentrations through $h_3$ and $h_4$, and can be obtained by~\cite{rf:5,rf:14}
\begin{eqnarray}
c_{\rm{B}}&=&(1-\Delta)\frac{3u_4h_4}{m_4n_4(d+h_4)^4},\nonumber
\\&\Delta&=\frac{u_3}{u_4}\Bigl[1-(\frac{d+h_4}{d+h_3+h_4})^4\Bigr]\,\,\,.
\end{eqnarray}
It is suggested that the third sound velocity Eq.(5) needs some corrections when the submonolayer regime of $^3$He is considered~\cite{rf:15}. We neglect these corrections here, because they does not play a crucial role in this paper.
\begin{figure}
\includegraphics[scale=0.33]{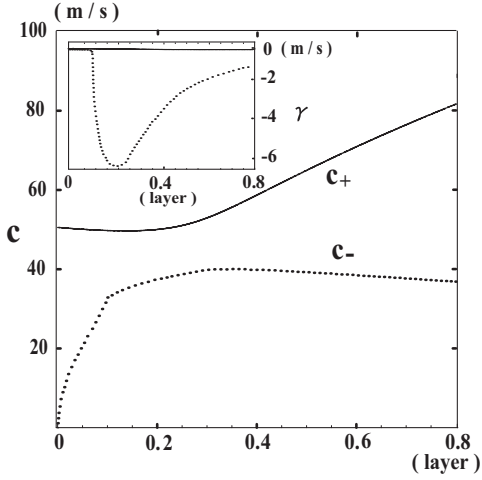}
\caption{\label{fig:epsart}  Real part and imaginary part (inset) of spectrum are plotted with $\omega/q = c+i\gamma$ as a function of concentration of $^3$He. Here, we fix the thickness of $^4$He films to $d=$25.3 $\mu$mol/m$^2$, $h_4=$25.8 $\mu$mol/m$^2$, and graphite sheet is modeled as a substrate.}
\end{figure}

Similar model has been studied in detail to investigate the equation of state and correlation energy of $^3$He~\cite{rf:16,rf:17}, whereas we will focus on the low energy collective behaviors of mixture below. To investigate the spectrum of collective excitations, we shall consider a $2\times 2$ susceptibility matrix, which is in connection with area density fluctuations (thickness variation) in linear response theory, in which interactions are treated within random phase approximation (RPA)~\cite{rf:18,rf:19}. The unperturbed susceptibility and interaction matrix of the mixtures has the following form
\begin{equation}
\widehat{\chi }^{(0)}= 
\left(
\begin{array}{cc}
\chi_3 & 0 \\ 
0 & \chi_4
\end{array}
\right), 
\,\,\,\,\,\,
\widehat{V}= 
\left(
\begin{array}{cc}
v_3 & g_{\bf q} \\
g_{\bf q} & 0
\end{array}
\right)
.
\end{equation}
Here, $\chi_3$ and $\chi_4$ are the susceptibilities of 2D pure fermion and boson systems, and at zero temperature they are given as
\begin{eqnarray}
\chi_3({\bf q},\omega)&=&\frac{S}{(2\pi)^2}\int {\rm d}^2{\bf k}
\frac{n(\epsilon_{{\bf k}-{\bf q/2}})-n(\epsilon_{{\bf k}+{\bf q/2}})}
{\epsilon_{{\bf k}-{\bf q/2}}-\epsilon_{{\bf k}+{\bf q/2}}+\hbar \omega+i0^+}\nonumber
\\&=&-\frac{N(0)}{2}(1+\frac{\frac{i\omega}{v_{\rm{F}}q}}{\sqrt{1-(\frac{\omega}{v_{\rm{F}}q})^2}})
\end{eqnarray}
and
\begin{eqnarray}
\chi_4({\bf q},\omega)&=&\frac{2\omega_{\bf q}/\hbar}
{\omega^2-\omega_{\bf q}^2 +i0^+} \,\,\,,
\end{eqnarray}
where $\epsilon_{\bf{k}}=\hbar^2{\bf{k}}^2/2m_3-\epsilon_{\rm{F}}$ with $\epsilon_{\rm{F}}$ being Fermi energy and $n(\epsilon_{\bf k})$, $v_{\rm{F}}$, $N(0)=m_3S/\pi \hbar^2$ are Fermi-Dirac distribution, Fermi velocity and density of states at Fermi surface of $^3$He, respectively. In the equations above, terms of order $q^4$ are neglected since we are interested in low energy excitations.

Within the RPA treatment, the susceptibility matrix of mixture films can be obtained as a solution of $\widehat{\chi }=\widehat{\chi }^{(0)}+\widehat{\chi }^{(0)}\widehat{V}\widehat{\chi }$~\cite{rf:19}, which is
\begin{eqnarray}
\widehat{\chi }_{\rm{RPA}}
&=&\frac{1}{1-v_3\chi_3-2g_{\bf q}^2\chi_3\chi_4}\nonumber
\\& &\times\,\,\,\,
\left(
\begin{array}{cc}
\chi_3 & g_{\bf q}\chi_3\chi_4 \\
g_{\bf q}\chi_3\chi_4 & \chi_4(1-v_3\chi_3)
\end{array}
\right) \,\,\,,
\end{eqnarray}
where the factor $2$ in the last term of the denominator comes as a result of summing over the spin indices. If $g_{\bf q}= 0$, it can be easily seen that this susceptibility matrix becomes diagonal whose elements are usual RPA susceptibility, $\chi_3/(1-v_3\chi_3)$, and $\chi_4$. Therefore, two poles of this matrix correspond to the spectrum of two eigenmodes; one is zero sound branch of $^3$He due to the nonlinearity of van der Waals potential, and the other is third sound branch of $^4$He. 
\begin{figure}
\includegraphics[scale=0.3]{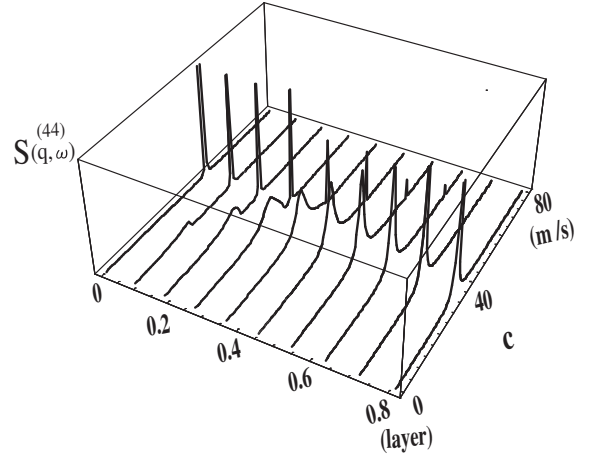}
\caption{\label{fig:epsart} Dynamical structure function S$^{(44)}(q,\omega)$ for a fixed wave number $q$ is plotted as a function of concentration of $^3$He and velocity of excitation. Third sound evolves to lower branch and spectral weight S$^{(44)}(q,\omega)$ is broadened as concentration of $^3$He quasiparticle increases. Parameters $u_i$, $d$ and $h_4$ are the same as the Fig.1.}
\end{figure}
For the third sound phonon, this RPA treatment is equivalent to solving Dyson's equation in Migdal approximation~\cite{rf:20}, where effective self-energy of third sound is $2g_{\bf q}^2\chi_3/(1-v_3\chi_3)$. At zero temperature and long wave length limit, dispersion relation and damping rate are calculated numerically and the results are shown in Fig.1. Here, we employs the parameters $u_i$ and $d$, modeling the double layer system on the flat surface of graphite~\cite{rf:3}. For the films, we fix the thickness of $^4$He films, and alter the $^3$He concentration from a small fraction of the monolayer to one layer (a monolayer corresponds to the density 10.6 $\mu$mol/m$^2$ for $^3$He and 12.9 $\mu$mol/m$^2$ for $^3$He~\cite{rf:8}). One can see that these two branches exhibit a level repulsion characteristic of reactively coupled oscillators. Hence, the two branch are well-hybridized and show the clear level splitting around the region $h_3 \approx 0.3$ layer ($\sim$ 3.2 $\mu$mol/m$^2$), where $c_{\rm{B}}$ crosses the bare zero sound velocity $c_{\rm{F}}=\frac{1+\tilde{v} /2}{\sqrt{1+\tilde{v} }}v_{\rm{F}}$ with $\tilde{v}=N(0)v_3$. As the concentration of $^3$He particles increases further, one can see from the dynamical structure function of $^4$He, S$^{(44)}(\bf{q},\omega)$, defined as
\begin{eqnarray}
\widehat{S }({\bf q},\omega )=-\frac{1}{\pi } \rm{Im} \,\widehat{\chi }({\bf q},\omega )=
\left(
\begin{array}{cc}
S^{(33)} & S^{(34)} \\
S^{(43)} & S^{(44)}
\end{array}
\right),
\end{eqnarray}
that third sound spectrum evolves continuously from the upper branch ($+$) into the lower one ($-$), and gets into the particle-hole continuum of the $^3$He. This situation results in the strong damping of bosonic third sound by resonantly absorbing $^3$He particle-hole pair. This behavior of the spectrum has not been verified~\cite{rf:15,rf:21}. To our knowledge, there is no experimental result, investigating carefully around the region where velocities of two components are close to each other.

As for the damping rate, the beginning of damping of hybridized mode can be seen in discontinuous concentration dependence of sound velocity in lower branch. Unlike conventional electron-phonon systems in metals, the ratio of sound velocities of two components is of order unity; therefore the effect of this damping is not small and has a possibility to be observed (see inset of Fig.1). Indeed, imaginary part of the spectrum can be easily estimated to have a behavior $\sim -\bigl[ \lambda/(2+\tilde{v}) \bigr] \bigl[ c_{\rm{B}}/\sqrt{v_{\rm{F}}^2-c_{\rm{B}}^2}\bigr](c_{\rm{B}}q)$ at high concentration regime of $^3$He, i.e., $c_{\rm{B}} < v_{\rm{F}}$. Here, $\lambda=-N(0)g_{\bf q}^2\chi_4({\bf q},0)>0$ is a dimensionless phonon exchange coupling constant, which is determined by Fermi energy and van der Waals potential~\cite{rf:5}. Around the region $c_{\rm{B}} \approx v_{\rm{F}}$, it has a strong peak which amounts to about 15 ${\tt\char"25}$ of the spectrum. In three dimensional (3D) case, similar level repulsion and the damping of bosonic mode are discussed by Yip with dilute boson-fermion mixture gasses of alkali atoms, such as $^{39}$K-$^{40}$K and $^{6}$Li-$^7$Li, in mind~\cite{rf:18}. In his system, fermions are spin polarized because of magnetic trapping and fermion-fermion interaction is not included, hence zero sound, due to phonon exchange interaction only, is discussed.

Having established the existence of the level repulsion and the damping within the RPA treatment, we will discuss effect of vertex function for $^3$He-$^4$He exchange interaction in more detail. Unlike conventional metals, ``Debye energy" is comparable or greater than $\epsilon_{\rm{F}}$ in our system. This situation opens the possibility of scenario in which nonadiabatic effects are relevant and new qualitative phenomena arise from vertex corrections. Here, we treat the vertex function in ladder approximation~\cite{rf:22}, i.e., as a solution of Bethe-Salpeter equation 
\begin{eqnarray}
\Gamma ({\bf q},\it{i}\omega_m) &=& g_{\bf q}
-k_{\rm{B}} T\sum_{\it{i} k_n}\sum_{\bf k} g_{{\bf k}-{\bf p}}^2\nonumber
\\&\times& \chi_4({\bf p}-{\bf k},\it{i}p_{l}-\it{i}k_n)
G({\bf k},\it{i}k_n)\nonumber
\\& &\times G({\bf k}-{\bf q},\it{i}k_n-\it{i}\omega_m)
\Gamma ({\bf q},\it{i}\omega_m) ,
\end{eqnarray}
where $G({\bf k},\it{i}k_m)$ is a bare single-particle propagator of $^3$He quasiparticles with $ik_n$, $i\omega_m$ being fermionic and bosonic Matsubara frequency. We assume that the vertex function depends only on the energy-momentum transfer. Here, external fermion frequency $ip_l$ is set to zero after analytic continuation. This approximation may be justified when low energy particle-hole excitations give dominant contribution.  As for the external wave vector ${\bf p}$, we introduce $q_{\rm {c}}$ as a cutoff of wave number transfer $|\bf{p}-\bf{k}|$~\cite{rf:23}, which is of order of inverse of coherence length of $^4$He films. At zero temperature and long wave length limit, this equation is solved analytically and the vertex function has the following form
\begin{eqnarray}
\Gamma ({\bf q},\omega) &=& \frac{g_{\bf q}}{1+\frac{\lambda}{2} \Bigl(C({\bf q},\omega)+\frac{\frac{i\omega}{v_{\rm{F}}q}}{\sqrt{1-(\frac{\omega}{v_{\rm{F}}q})^2}} \Bigr) }\,\,\,,
\\
C({\bf q},\omega) &=& \frac{1}{2}\Bigl(\frac{1}{\sqrt{1-(\frac{v_{\rm{F}} q}{c_{\rm{B}} q_{\rm {c}}-\omega})^2}}+\frac{1}{\sqrt{1-(\frac{v_{\rm{F}} q}{c_{\rm{B}} q_{\rm {c}}+\omega})^2}} \Bigr).\nonumber
\\
\end{eqnarray}
Here, we neglect polarization contribution of the second term of right hand side in Eq.(11), which has the same dynamical ($\omega\rightarrow 0$, ${\bf q}=0$) and static limit ($\omega =0$, ${\bf q}\rightarrow0$), because it gives only quantitative difference in the denominator of vertex function. Obviously, Migdal approximation breaks down qualitatively in the region $\omega\approx v_{\rm{F}}q$ and the vertex function diverges to $\pm \infty$, as $\omega \rightarrow \omega_{\rm {C}}$ from above and below respectively. Here, $\omega_{\rm {C}}$ is obtained as
\begin{equation}
\omega_{\rm {C}} \simeq \frac{1+\lambda/2}{\sqrt{1+\lambda }}
\Bigl(v_{\rm{F}}q-\frac{\lambda(1+\lambda^2/2)}{8(1+\lambda/2)^2}\frac{(v_{\rm{F}}q)^3}{(c_{\rm{B}} q_{\rm {c}})^2}\Bigr) .
\end{equation}
This means that dynamical phonon-mediated $^3$He-$^3$He interaction becomes strongly enhanced and even changes its sign. Its behavior resembles Feshbach resonance in alkali atom gasses~\cite{rf:24}, but this frequency dependent ``potential" cannot be used in a Hamiltonian formalism like a scattering length of alkali atoms since the strong frequency dependence of interaction imposes one to take the strong retardation effect into account~\cite{rf:22}. We expect this break down of perturbation theory not to lead to the violation of Fermi liquid theory. Indeed, the pole provides only a subdominant correction to imaginary part of single-particle self-energy that behaves as Im$\Sigma^p(k_F,\omega)\sim \lambda^3 \epsilon_F(\hbar \omega/\epsilon_F)^2$ for $\omega \rightarrow 0^+$ (subscript $p$ denotes the pole contribution). Moreover, nonzero quasiparticle residue $Z$ at the Fermi surface can be derived with use of Kramers-Kronig relation. Thus the quasiparticle is well-defined, even with the singularity in vertex function.

We shall investigate the effect of this phonon exchange vertex function to the spectrum of excitations. The singularity of $\Gamma({\bf q},\omega)$ in long wave length limit implies the existence of low-lying excitation which obeys Bose-Einstein statistics. We interpret the pole as a excitation energy of an additional collective mode. Physically, this third branch corresponds to a combined mode of $^3$He particle-hole pair and third sound phonon. It may be considered as the pole of $D_{\rm {C}}(\bf q,\omega)$, which is Fourier transformation of $-\sum_{\sigma} \langle T_t(\phi_{\bf q}(t)\rho_{{-\bf q},\sigma}(0)+\rho_{{\bf q},\sigma}(t)\phi_{-\bf q}(0)) \rangle$~\cite{rf:20}. Here, $T_t$ is usual Wick's time ordering operator, $\phi_{\bf q}(t)$ is field operator of third sound phonon, and $\langle \cdots \rangle$ denotes the average for the ground state of mixture. It should be noted that this ``susceptibility" contains multi-phonon processes, thus relation with area density fluctuations is no longer linear. The third spectrum obtained by numerical calculation lies above particle-hole continuum and between upper and lower branch. Furthermore, one can see from dynamical structure function $S^{C}({\bf q}, \omega)=-1/\pi{\rm{Im}}D_C$ that the combined mode branch has considerable spectral weight except the region where zero sound and third sound are well-hybridized. The behavior of this spectral weight causes the resonant nature of $S^{C}({\bf q}, \omega)$ at $\omega \approx c_{\rm{B}}q$. Indeed, the combined mode is completely suppressed as $\omega_{\rm {C}} \rightarrow c_{\rm{B}}q$. Physically, this behavior causes the mode-mode repulsion~\cite{rf:18}.
\begin{figure}
\includegraphics[scale=0.35]{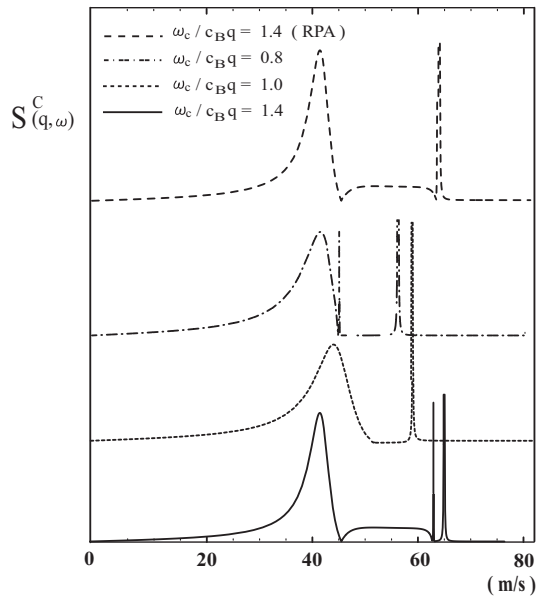}
\caption{\label{fig:epsart} Dynamical structure function of the combined mode, S$^{C}$$(q,\omega)$, is plotted for three parameter regimes as a function of $\omega/q$. One can see that there exists a sharp spectral weight between third sound and zero sound, which does not occur within RPA treatment. At $\omega_{\rm {C}}/c_{\rm{B}}q=1.4$, spectral weight of third branch is found to be about 30 ${\tt\char"25}$ of all. At $\omega_{\rm {C}}/c_{\rm{B}}q=1.4$, spectral weight of third branch is found to be about 30 ${\tt\char"25}$ of all. At the region $\omega_{\rm {C}} \approx c_{\rm{B}}q$, the combined mode is strongly suppressed by mode-mode repulsion effect.}
\end{figure}

We also investigate the correction of the combined mode to Landau Fermi liquid parameter. At zero temperature, it is obtained as $f_{\sigma ,\sigma^{'}}(\theta)= \delta \Sigma_{\sigma}({\bf k},\epsilon_{\bf k})/ \delta n_{\sigma^{'}}(\epsilon_{\bf  k^{'}})$ with ${\bf k} \cdot {\bf k^{'}}=k_{\rm{F}}^2 \rm{cos} \theta$~\cite{rf:25}. Since the combined mode does not contribute to $f_{\uparrow \downarrow}(\theta)$, we can obtain the Landau {\it f} function as follows~\cite{rf:25}
\begin{eqnarray}
f_{\uparrow \uparrow}(\theta)=f_{\uparrow \downarrow}(\theta)-{\rm P}.\int \frac{{\rm d} s}{\pi} \frac{\lambda }{s} \frac{{\rm Im} \Gamma ({\bf k}-{\bf k'},s)}{g_{{\bf k}-{\bf k'}}}.
\end{eqnarray}
Here, {\rm P.} denotes the principal value of the integral. The pole contribution, i.e. the collective mode contribution, comes from delta function part of Im$\Gamma$, and is obtained as
\begin{eqnarray}
N(0)(f_{\uparrow \uparrow}^p(\theta)-f_{\uparrow \downarrow}^p(\theta))\approx -\frac{\lambda^3}{16} \Bigl(1+\frac{\lambda }{2}\frac{(v_{\rm{F}}k_{\rm{F}})^2}{(c_{\rm{B}} q_{\rm {c}})^2}\rm{sin}^2(\frac{\theta}{2}) \Bigr),\nonumber
\\
\end{eqnarray}
In addition, contribution from the particle-hole continuum of vertex function can be obtained from the real part of $\Gamma$ with use of Kramers-Kronig relation, and has the following form
\begin{eqnarray}
N(0)(f_{\uparrow \uparrow}^{ph}(\theta)-f_{\uparrow \downarrow}^{ph}(\theta))\approx \frac{-\lambda}{1+\frac{\lambda}{2}C(2k_{\rm{F}}\rm{sin}(\frac{\theta}{2}),0)}.
\end{eqnarray}
These contributions are not singular. Additionally, the pole gives rise an additional factor $\lambda ^2$ to correction for Landau {\it f} function. Therefore, third branch contribution may become important at strong coupling region.

As for the observability, this third branch may be observed under the strong van der Waals potential, since it is inherently related to the interaction. Therefore experiment with low-thickness superfluid $^4$He films and the substrate which has strong van der Waals potential, such as gold, is desired. Recent experiment makes it possible to evaluate the Landau Fermi liquid parameter as a function of $^3$He concentration~\cite{rf:4}, therefore the third branch contribution Eq.(16) has a possibility to be observed at the low-concentration regime of $^3$He. Further, it should be noted that there is no restriction in system components in our method, therefore these discussions can be easily applied to quasi-2D alkali atom gases in which phase separation does not occur. Regarding existence of the third branch, alkali gases system may be better suited for experimental verification because fermion-boson interaction can be almost freely changed as well as other system parameters. In such a case, zero sound with higher spin channel may be taken into account~\cite{rf:26}.

In conclusion, we have investigated the effects of phonon exchange process to the collective excitations of $^3$He and $^4$He at absolute zero. Two remarkable features have been found. One is the level repulsion between zero sound and third sound due to $^3$He-$^4$He interaction. This repulsion becomes remarkable at region where sound velocities of two components are close to each other. This situation results in hybridization of two eigenmodes and finite level splitting in the spectrum. Furthermore, damping of third sound, resonant decay into particle-hole pair excitation, is also shown to be significant when the third sound velocity is smaller than $v_{\rm{F}}$. The other is that Migdal's theorem breaks down when nonadiabatic effect of $^3$He-$^4$He interaction is considered. A third branch comes in; this may be interpreted as a combined mode of particle-hole pair and third sound quanta. The concentration dependence of dynamical structure factor and the correction to Landau Fermi liquid parameter of this collective excitation are also discussed. 

In the discussions above, we have focused on low energy collective behavior and we have not considered two-body scattering problem. It is expected that interaction between $^3$He and $^4$He induces the effective $^3$He-$^3$He attraction and leads to spin-singlet and neutral superfluid formation~\cite{rf:5,rf:6,rf:7,rf:8}. We consider in future work the possibility that nonadiabatic phonon exchange may either drive the system into a superfluid transition or polaron formation.

The authors would like to thank M. Nishida, S. Tsuchiya, B. H. Valtan, and A. Morales for useful discussions.


\end{document}